\RequirePackage{ifpdf}
\ifpdf 
\documentclass[pdftex]{sigma}
\else
\documentclass{sigma}
\fi

\newcommand{\starcom}[2]{[#1\stackrel{\star}{,}#2]}
\def\vol{\mathrm{vol}}

\numberwithin{equation}{section}

\begin{document}

\allowdisplaybreaks

\renewcommand{\thefootnote}{$\star$}

\renewcommand{\PaperNumber}{061}

\FirstPageHeading

\ShortArticleName{Field Theory on Curved Noncommutative Spacetimes}

\ArticleName{Field Theory on Curved Noncommutative Spacetimes\footnote{This paper is a
contribution to the Special Issue ``Noncommutative Spaces and Fields''. The
full collection is available at
\href{http://www.emis.de/journals/SIGMA/noncommutative.html}{http://www.emis.de/journals/SIGMA/noncommutative.html}}}

\Author{Alexander SCHENKEL and Christoph F.~UHLEMANN}

\AuthorNameForHeading{A.~Schenkel and C.F.~Uhlemann}

\Address{Institut f\"ur Theoretische Physik und Astrophysik,     Universit\"at W\"urzburg,\\
  Am Hubland, 97074 W\"urzburg, Germany}
\Email{\href{mailto:aschenkel@physik.uni-wuerzburg.de}{aschenkel@physik.uni-wuerzburg.de}, \href{mailto:uhlemann@physik.uni-wuerzburg.de}{uhlemann@physik.uni-wuerzburg.de}}

\ArticleDates{Received March 17, 2010, in f\/inal form July 14, 2010;  Published online August 03, 2010}

\Abstract{We study classical scalar f\/ield theories on noncommutative curved spacetimes. Following the approach
of Wess et al.~[{\it Classical Quantum Gravity} {\bf 22} (2005), 3511 and {\it Classical Quantum Gravity} {\bf 23} (2006), 1883],
we describe noncommutative spacetimes by using (Abelian) Drinfel'd
twists and the associated $\star$-products and $\star$-dif\/ferential geometry. In particular, we allow for position dependent
noncommutativity and do not restrict ourselves to the Moyal--Weyl deformation.
We construct action functionals for real scalar f\/ields on noncommutative curved spacetimes, and derive the corresponding
deformed wave equations. We provide explicit examples of deformed Klein--Gordon operators for noncommutative
Minkowski, de Sitter, Schwarzschild and Randall--Sundrum spacetimes, which solve the noncommutative
Einstein equations. We study the construction of deformed Green's functions and provide a diagrammatic approach for their
perturbative calculation. The leading noncommutative corrections to the Green's functions for our examples are derived.}


\Classification{81T75; 83C65; 53D55}

\renewcommand{\thefootnote}{\arabic{footnote}}
\setcounter{footnote}{0}

\section{Introduction}
Noncommutative (NC) geometry \cite{Connes:1994yd} is a very rich framework for modifying the kinematical structures of low-energy theories.
In this approach the ingredients for a classical description of spacetime (manifolds, vector bundles, \dots)
are generalized to suitable quantum objects (algebras, projective modules, \dots). For an introduction to
NC geometry see also \cite{Landi:1997sh}. Replacing classical spacetime by
NC spaces has been motivated from dif\/ferent perspectives. There are Gedanken experiments indicating
that the precise localization of an event in spacetime can induce NC~\cite{Doplicher}, as well as indications
showing that NC geometry can emerge from string theory~\cite{Seiberg:1999vs} and quantum gravity~\cite{quantumgrav}.
Another motivation for studying NC spacetimes is the hope that replacing all classical spaces (including spacetime) by appropriate quantum spaces
could improve the mathematical description of physics, e.g.~concerning the UV divergences in quantum f\/ield theory
or the curvature singularities in general relativity.

A natural possibility to describe NC gravity is to employ a NC metric f\/ield \cite{Aschieri:2005yw,Aschieri:2005zs,Kurkcuoglu:2006iw},
but there are also other well motivated approaches based on
hermitian metrics \cite{Chamseddine:1992yx,Chamseddine:2000zu} or gauge formulations with or without Seiberg--Witten maps
using dif\/ferent gauge groups  \cite{Chamseddine:2000si,Cardella:2002pb,Chamseddine:2003we,Banerjee:2007th,Aschieri:vb}.
See also~\cite{ncgravity} for a collection of approaches towards NC gravity.
In addition to the formulations which are deformations of the classical framework using metrics (or similar ingredients),
NC geometry also of\/fers a natural mechanism for emergent gravity within NC gauge theory and matrix
models \cite{Rivelles:2002ez,Yang,Steinacker}.

\looseness=1
In this work we follow the formulation proposed by Julius Wess and his group \cite{Aschieri:2005yw,Aschieri:2005zs}
to describe NC gravity and f\/ield theories. As ingredients we use
$\star$-products instead of abstract operator algebras. This approach is called deformation quantization \cite{Dito:2002dr}
and has the advantage that the quantum theory is formulated in terms of the classical objects, thus allowing us
to study deviations (perturbations) from the classical situation at every step.
 Obviously, formal deformation quantization has the disadvantage that we may miss interesting examples, where
NC is very strong. An interesting feature of the formulation \cite{Aschieri:2005yw,Aschieri:2005zs} is
that the NC spaces obey ``quantum symmetry'' properties, since the $\star$-products
are constructed by Drinfel'd twists \cite{Drinfeld}. This is an advantage compared to generic NC spaces,
since symmetries are an important guiding principle for constructing f\/ield theories, in particular gravity theories.

Recently, there has been considerable progress towards applications of the NC gravity theory of Wess  et al.~to
physical situations. Symmetry reduction, the basic tool for studying symmetric conf\/igurations in gravity,
has been investigated in \cite{Ohl:2008tw} in theories obeying quantum symmetries. Furthermore, pioneered by Schupp and
Solodukhin \cite{Schupp:2009pt}, exact solutions of the NC Einstein equations have been found in
\cite{Schupp:2009pt,Ohl:2009pv,Aschieri:2009qh,Asakawa:2009yb,Stern:2009id}, providing, in particular, explicit models
for NC cosmology and NC black hole physics. For a collection of other approaches towards
NC cosmology see \cite{NCcosmo}, and for NC black holes see \cite{NCblack}.

The natural next step is to consider (quantum) f\/ield theory on these recently obtained curved NC spacetimes
in order to make contact to physics, like e.g.~the cosmic microwave background or Hawking radiation.
(Quantum) f\/ield theory on NC spacetimes is a very active subject, see~\cite{NCQFT} for a collection of dif\/ferent approaches.
However, most of these studies are restricted to the Moyal--Weyl or $\kappa$-deformed Minkowski spacetime.
In~\cite{Ohl:2009qe} we attempt to f\/ill this gap and proposed a mathematical framework for quantum f\/ield theory (QFT)
on curved NC spacetimes within the algebraic approach to QFT \cite{Wald,baer}. We have shown that
for a large class of Drinfel'd twist deformations one can construct deformed algebras of observables
for a free real scalar f\/ield. However, the construction of suitable quantum states and the physical application of this formalism
still have to be investigated.

The purpose of this article is to review the formalism of \cite{Ohl:2009qe}  in a rather nontechnical language and apply it
 to study explicit examples of classical f\/ield theories on curved NC spacetimes.
 The outline of this paper is as follows:
In Section~\ref{sec:formalism} we review the NC geometry
from Drinfel'd twists, restricting ourselves to the class of Abelian (also called RJS-type) twists. Using these methods we show
in Section~\ref{sec:fieldtheory} how to construct deformed action functionals for scalar f\/ields
on NC curved spacetimes,
also allowing for position dependent NC. Additionally to the abstract geometric formulation
of~\cite{Ohl:2009qe} we also present the construction of deformed actions and equations of motion
using a local basis, which is the language typically used in the physics literature. In Section~\ref{sec:examples1}
we study examples of the deformed Klein--Gordon operators derived in Section \ref{sec:fieldtheory} using dif\/ferent
NC Minkowski spaces, de Sitter universes, a Schwarzschild black hole and a Randall--Sundrum spacetime.
We provide a convenient formalism to study NC corrections to the Green's functions of the deformed
equations of motion in Section~\ref{sec:greens} and apply it to examples in Section~\ref{sec:examples2}.
We conclude in Section~\ref{sec:conclusions}.

\section{NC geometry from Drinfel'd twists}\label{sec:formalism}

In this section we provide the required background on $\star$-products and NC geometry
from Drinfel'd twists.
We omit mathematical details as much as possible and use simple examples to explain the formalism.
Mathematical details on NC geometry (and gravity) from Drinfel'd twists can be found in
 \cite{Aschieri:2005yw,Aschieri:2005zs}. Furthermore, we frequently write expressions in local coordinates and use local bases
 of vector f\/ields and one-forms, as it is typically done in the physics literature.
For a global and coordinate independent formulation see \cite{Aschieri:2005zs}.

Instead of providing the abstract def\/inition of a $\star$-product, let us study the simple example
of the Moyal--Weyl product and emphasize the basic features. Assume spacetime to
be $\mathcal{M}=\mathbb{R}^D$. At this point we do not require a metric f\/ield on $\mathcal{M}$. Let
$h,k\in C^\infty(\mathcal{M})$ be two smooth, complex-valued functions on $\mathcal{M}$. We replace the classical
point-wise multiplication of functions by the Moyal--Weyl product
\begin{gather}
 \nonumber (h \star k)(x) := h(x) \exp\left(\frac{i\lambda}{2}\overleftarrow{\partial_\mu}\Theta^{\mu\nu}\overrightarrow{\partial_\nu}\right) k(x)\\
\phantom{(h \star k)(x)}{}
= \sum\limits_{n=0}^{\infty} \left(\frac{i \lambda}{2}\right)^n\frac{1}{n!} \Theta^{\mu_1\nu_1}\cdots\Theta^{\mu_n\nu_n} \bigl(\partial_{\mu_1}\cdots\partial_{\mu_n} h(x)\bigr)  \bigl(\partial_{\nu_1}\cdots\partial_{\nu_n} k(x)\bigr)  ,\label{eqn:moyalproduct}
\end{gather}
where $\Theta^{\mu\nu}$ is a constant and antisymmetric matrix, $\lambda$ is the deformation parameter and
$\partial_\mu$ are the partial derivatives with respect to the coordinate directions $x^\mu$.
It can be checked easily that the $\star$-product is associative, i.e.~that $h\star (k\star l)=(h\star k)\star l$
for all $h,k,l\in C^\infty(\mathcal{M})$, but noncommutative $h\star k \neq k\star h$.
Applying the $\star$-product to the coordinate functions $x^\mu$ we obtain the commutation relations
\begin{gather}
\label{eqn:moyalcom}
 \starcom{x^\mu}{x^\nu}:= x^\mu\star x^\nu -x^\nu\star x^\mu = i\lambda \Theta^{\mu\nu}.
\end{gather}
Thus, using the Moyal--Weyl product, we obtain a NC space similar to the NC
phasespace of quantum mechanics. However, in this case spacetime itself is NC.

Before generalizing the $\star$-product (\ref{eqn:moyalproduct}) to include position dependent NC,
i.e.~position dependent commutation relations (\ref{eqn:moyalcom}), we note that the
$\star$-product can be written using the bi-dif\/ferential operator
\begin{gather}
\label{eqn:moyaltwist}
 \mathcal{F}^{-1}_{\text{MW}} := \exp\left( \frac{i\lambda}{2}\Theta^{\mu\nu}\partial_\mu \otimes\partial_\nu \right),
\end{gather}
by f\/irst applying $\mathcal{F}^{-1}_\text{MW}$ to $h\otimes k$ and then multiplying the result using the point-wise
multiplication. The object $\mathcal{F}_{\text{MW}}$ is a particular example of a Drinfel'd twist \cite{Drinfeld}.

We now generalize the Moyal--Weyl product (\ref{eqn:moyalproduct}) to a class of (possibly position dependent) $\star$-products
on a general manifold $\mathcal{M}$. Consider a set of commuting (w.r.t.~the Lie bracket) vector f\/ields $\lbrace X_\alpha\rbrace$,
where $\alpha$ is a label, {\it not} a spacetime index. In local coordinates we have $X_\alpha = X_\alpha^\mu(x)\partial_\mu$.
The vector f\/ields $X_\alpha$ are def\/ined to act on functions as (Lie) derivatives, i.e.~in local coordinates we have
$X_\alpha h = X_\alpha^\mu(x)\partial_\mu h(x) $. Using $\lbrace X_\alpha\rbrace$ and a constant and antisymmetric matrix
$\Theta^{\alpha\beta}$ we can def\/ine the following $\star$-product
\begin{gather}
\label{eqn:rjsproduct}
(h\star k)(x):= h(x)\exp\left(\frac{i\lambda}{2}\overleftarrow{X_\alpha}\Theta^{\alpha\beta}\overrightarrow{X_\beta} \right) k(x) .
\end{gather}
Associativity of this product is easily shown using $[X_\alpha,X_\beta]=0$.
Furthermore, we can analogously to (\ref{eqn:moyaltwist}) associate a Drinfel'd twist to the $\star$-product (\ref{eqn:rjsproduct}),
namely
\begin{gather}
\label{eqn:rjstwist}
 \mathcal{F}^{-1} := \exp\left(\frac{i\lambda}{2}\Theta^{\alpha\beta}X_\alpha \otimes X_\beta \right).
\end{gather}
These twists are called Abelian or of Reshetikhin--Jambor--Sykora (RJS) type
\cite{Reshetikhin:1990ep,Jambor:2004kc}. In the following we restrict
ourselves to real vector f\/ields $X_\alpha$ (i.e.~unitary and real twists),  leading to
hermitian $\star$-products $(h\star k)^\ast = k^\ast \star h^\ast$. Furthermore, we can without loss of
generality assume $\Theta^{\alpha\beta}$ to be of the canonical (Darboux) form
\begin{gather*}
 \Theta^{\alpha\beta} = \begin{pmatrix}
         0 & 1 & 0 & 0 &\cdots \\
	-1 & 0 & 0 & 0 &\cdots\\
	 0 & 0 & 0 & 1 &\cdots\\
	 0 & 0 & -1 & 0 & \cdots \\
	 \vdots & \vdots & \vdots & \vdots & \ddots
        \end{pmatrix}.
\end{gather*}

Let us brief\/ly discuss two simple examples of non-Moyal--Weyl twists and $\star$-products.
Let $\mathcal{M}=\mathbb{R}^{D}$.
Consider $X_1=\partial_t$ and $X_2 = x^i \partial_i$, where $t$, $x^i$, $i=1,\dots,D-1,$ are global coordinates on $\mathbb{R}^{D}$. It is easy to check
that $[X_1,X_2]=0$, thus we obtain a $\star$-product of RJS type. The commutation relations of the
coordinate functions are given by $\starcom{t}{x^i} = i\lambda x^i$ and $\starcom{x^i}{x^j}=0$.
These are the commutation relations of a $\kappa$-deformed spacetime~\cite{kappa}.
For the second example consider $\mathcal{M}=\mathbb{R}^2$ and $X_1=x \partial_x$, $X_2=y\partial_y$, where $x$, $y$ are global coordinates.
We obtain the commutation relation of the quantum plane $x\star  y= q  y\star x$, where $q=e^{i\lambda}$.

Since our aim is to describe f\/ield theory on the NC spacetimes $(C^\infty(\mathcal{M}),\star)$\footnote{The mathematically precise def\/inition of the algebra is $(C^\infty(\mathcal{M})[[\lambda]],\star)$.
We suppress the brackets $[[\lambda]]$ indicating formal power series for better readability.},
we have to introduce a few more ingredients, such as derivatives, metrics and integrals, to the NC setting.
The notion of a derivative is described by
a dif\/ferential calculus over the algebra $(C^\infty(\mathcal{M}),\star)$. In our particular example of deformations using
Drinfel'd twists, the dif\/ferential calculus is given by the dif\/ferential forms on the manifold $\Omega^\bullet$, the deformed wedge product $\wedge_\star$ and the undeformed exterior derivative $d$. The deformed
wedge product is def\/ined by $\omega\wedge_\star \omega^\prime := \wedge \bigl( \mathcal{F}^{-1} \omega\otimes\omega^\prime\bigr)$, i.e.~f\/irst acting with the inverse twist via the Lie derivative on $\omega\otimes\omega^\prime$
and then multiplying  by the standard wedge product. We also give the explicit expression for
 $\wedge_\star$ in presence of an RJS twist~(\ref{eqn:rjstwist})
\begin{gather}\label{eqn:star-wedge}
\omega \wedge_\star\omega^\prime = \sum\limits_{n=0}^{\infty} \left(\frac{i \lambda}{2}\right)^n\frac{1}{n!} \Theta^{\alpha_1\beta_1}\cdots\Theta^{\alpha_n\beta_n} \bigl(\mathcal{L}_{X_{\alpha_1}}\cdots\mathcal{L}_{X_{\alpha_n}}\omega\bigr) \wedge \bigl(\mathcal{L}_{X_{\beta_1}}\cdots\mathcal{L}_{X_{\beta_n}}\omega^\prime\bigr),
\end{gather}
where $\mathcal{L}_X$ denotes the Lie derivative along the vector f\/ield $X$. Using that Lie derivatives commute with the exterior
derivative $d$, one f\/inds that
\[
d(\omega\wedge_\star\omega^\prime)=(d\omega)\wedge_\star \omega^\prime + (-1)^{\text{deg}(\omega)} \omega\wedge_\star (d\omega^\prime),
\]
for all dif\/ferential forms $\omega,\omega^\prime\in\Omega^\bullet$.

Having introduced vector f\/ields $\Xi$ and one-forms $\Omega$ (co-vector f\/ields), we can consider the pairing (index contraction) in the
 NC setting. We def\/ine in the canonical way
 $\langle \omega, v\rangle_\star := \langle \cdot , \cdot  \rangle \bigl(\mathcal{F}^{-1}\omega\otimes v\bigr)$,
 where $\langle\cdot,\cdot\rangle\bigl(\omega\otimes v\bigr)=\langle \omega,v\rangle$ is the commutative pairing,
  given in a local coordinate basis $v=v^\mu\partial_\mu$, $\omega=dx^\mu \omega_\mu$ by
  $\langle\omega,v \rangle=\omega_\mu v^\mu$. Again, we provide the explicit expression
\begin{gather*}
  \langle\omega, v \rangle_\star = \sum\limits_{n=0}^{\infty} \left(\frac{i \lambda}{2}\right)^n\frac{1}{n!} \Theta^{\alpha_1\beta_1}\cdots\Theta^{\alpha_n\beta_n}\langle \mathcal{L}_{X_{\alpha_1}}\cdots\mathcal{L}_{X_{\alpha_n}}\omega,  \mathcal{L}_{X_{\beta_1}}\cdots\mathcal{L}_{X_{\beta_n}}v\rangle.
\end{gather*}
The pairing $\langle v, \omega \rangle_\star$ with the vector f\/ield on the left is def\/ined
analogously.

Another ingredient for formulating NC f\/ield theories is a background spacetime metric f\/ield~$g$.
To def\/ine it we introduce the $\star$-tensor product $\otimes_\star$, which
can be constructed in the canonical way~\cite{Aschieri:2005zs}
by f\/irst acting with the twist~(\ref{eqn:rjstwist}) and then applying the usual tensor product.
For the RJS-twist~(\ref{eqn:rjstwist}) we have the following explicit form
\[
\tau\otimes_\star\tau^\prime := \sum\limits_{n=0}^{\infty} \left(\frac{i \lambda}{2}\right)^n\frac{1}{n!} \Theta^{\alpha_1\beta_1}\cdots\Theta^{\alpha_n\beta_n}\bigl( \mathcal{L}_{X_{\alpha_1}}\cdots\mathcal{L}_{X_{\alpha_n}}\tau\bigr)  \otimes \bigl(\mathcal{L}_{X_{\beta_1}}\cdots\mathcal{L}_{X_{\beta_n}}\tau^\prime\bigr) ,
\]
where $\tau$, $\tau^\prime$ are either vector f\/ields or one-forms.
The extension of $\otimes_\star$ to higher tensor f\/ields is straightforward and one obtains an associative tensor
algebra $(\mathcal{T},\otimes_\star)$ generated by~$\Xi$ and~$\Omega$~\cite{Aschieri:2005zs}.
We use a minor generalization of the formalism of~\cite{Aschieri:2005zs} and def\/ine the metric $g\in\Omega\otimes_\star\Omega$ to
be a~hermitian and nondegenerate tensor.  Note that the case
of real and symmetric metric f\/ields~$g$ is included in our def\/inition, thus every classical metric f\/ield is also a NC
metric. The inverse metric f\/ield
$g^{-1}=g^{-1\alpha}\otimes_\star g^{-1}_\alpha\in \Xi\otimes_\star\Xi$ (sum over $\alpha$ understood)
is also nondegenerate and hermitian.  We use the inverse metric f\/ield to
contract two one-forms $\omega,\omega^\prime\in\Omega$ to obtain a~scalar function.
This contraction is used later to def\/ine a kinetic term in the action. The metric contraction is called hermitian
structure and is def\/ined by
\begin{gather*}
H_\star(\omega,\omega^\prime) := \langle\langle\omega^\ast,g^{-1}\rangle_\star,\omega^\prime\rangle_\star := \langle \omega^\ast,g^{-1\alpha}\rangle_\star \star \langle g^{-1}_\alpha,\omega^\prime  \rangle_\star ,
\end{gather*}
where $\ast$ denotes conjugation on one-forms.
The hermitian structure is nondegenerate, hermitian and fulf\/ils $\star$-sesquilinearity
\begin{subequations}
\label{eqn:hermitianproperties}
\begin{gather}
H_\star(\omega,\omega^\prime)=0\quad \text{for all~}\omega^\prime\quad\Longleftrightarrow\quad \omega=0,\\
H_\star(\omega,\omega^\prime)^\ast = H_\star(\omega^\prime,\omega), \label{eqn:hermitianproperties2}\\
H_\star(\omega\star h,\omega^\prime\star k)=h^\ast\star H_\star(\omega,\omega^\prime)\star k,\label{eqn:hermitianproperties3}
\end{gather}
\end{subequations}
for all $\omega,\omega^\prime\in\Omega$ and $h,k\in C^\infty(\mathcal{M})$.

The last ingredient we require in the following is the integral $\int$ over the manifold $\mathcal{M}$.
In a geometrical language, the integral over spacetime associates to a top-form, i.e.~a dif\/ferential form $\tau$
with $\text{deg}(\tau) = \text{dim}(\mathcal{M})$, a number $\int \tau\in\mathbb{C}$.  The integral is evaluated locally
using the coordinate charts. Note that, since NC and commutative dif\/ferential forms are as vector spaces
the same (up to the formal power series), we can use the commutative integral also in the NC case.
Furthermore, one explicitly observes using (\ref{eqn:star-wedge}) and integration by parts that
\begin{gather}
\label{eqn:gradedcyclicity}
\int \omega\wedge_\star\omega^\prime = (-1)^{\text{deg}(\omega) \text{deg}(\omega^\prime)}\int\omega^\prime\wedge_\star\omega=\int\omega\wedge\omega^\prime,
\end{gather}
for all $\omega,\omega^\prime\in\Omega^\bullet$ with $\text{deg}(\omega)+\text{deg}(\omega^\prime)=\dim (\mathcal{M})$
and $\text{supp}(\omega)\cap\text{supp}(\omega^\prime)$ compact (in order to avoid boundary terms).
This property, called graded cyclicity, simplif\/ies the derivation of the equations of motion.
Note that, although graded cyclicity is fulf\/illed for all RJS-twists~(\ref{eqn:rjstwist}),
for the most general twists it is not.
Whether graded cyclicity is necessary, or just convenient, for the construction of the NC scalar f\/ield theory presented in the next sections
is an interesting question left for the future.

\section{NC scalar f\/ield action and equation of motion}\label{sec:fieldtheory}

The formulation of classical and quantum f\/ield theories on NC spacetimes has been a very active subject
over the last few years, see e.g.~\cite{NCQFT}. Most of these approaches focus on free or interacting QFTs
 on the Moyal--Weyl deformed  or $\kappa$-deformed Minkowski spacetime. In order to address physical
applications like QFT in a NC early universe or on a NC black hole background,
a formulation which can be extended to curved spaces has to be developed. For globally hyperbolic spacetimes deformed
by a large class of Drinfel'd twists (in particular including the RJS twists~(\ref{eqn:rjstwist})) we gave a proposal~\cite{Ohl:2009qe}
of how to construct the QFT of a free real scalar f\/ield using the algebraic approach to QFT~\cite{Wald, baer}. In our approach we have treated the deformation parameter $\lambda$ as a formal parameter, thus obtaining
a perturbative framework, which nevertheless could be solved formally to all orders in the deformation parameter.
The advantage of our approach is that we can apply it to study free quantum f\/ields on curved spacetimes with an in
general position dependent NC. The obvious disadvantage is that we treat the deformation parameter
as a formal parameter, thus we might not be able to capture nonperturbative NC ef\/fects, such as a
possible causality violation.

In the f\/irst part of this section we use the formalism of~\cite{Ohl:2009qe} to def\/ine actions and derive equations of motion
for a real scalar f\/ield on curved NC spacetimes using a geometric language. In the second part, we use local bases of
vector f\/ields and one-forms in order to rewrite the formalism using ``indices''.
This is important for constructing examples lateron, since, despite the elegance of the geometrical approach,
practical calculations are performed in a local basis.

\subsection{Basis free formulation}

We start by showing how to construct an action for a real scalar f\/ield $\Phi$ on NC curved spacetimes.
We are in particular interested in the deformation of the ``canonical action''
\[
S=-\frac{1}{2}\int \left(\partial_\mu\Phi  g^{\mu\nu}  \partial_\nu\Phi +m^2 \Phi^2 \right)\vol_g,
\]
where $\vol_g=\sqrt{\vert g\vert} d^Dx$ is the metric volume element (a $D$-form) and $D$ is the dimension of spacetime.
Using the tools of Section~\ref{sec:formalism} we can deform this action leading to the global expression
\begin{gather}
\label{eqn:scalaraction}
S_\star := -\frac{1}{2}\int \left(H_\star(d\Phi,d\Phi) +m^2 \Phi\star\Phi \right)\star\vol_\star,
\end{gather}
where $\vol_\star$  is a nondegenerate and real $D$-form\footnote{For general twist deformations there is, to our knowledge, no $\star$-covariant construction principle for a metric volume form.
One reasonable choice is
$\vol_\star=\sqrt{\vert \text{det} g_{\mu\nu}\vert}\frac{1}{D!}\epsilon_{\mu_1\dots \mu_D}dx^{\mu_1}\wedge\dots \wedge dx^{\mu_D}$,
constructed from the expression for $g$ in the commutative basis $g=g_{\mu\nu} dx^\mu\otimes dx^\nu$.
It is nondegenerate, real and has the correct classical limit.
In this section we keep $\vol_\star$ general, demanding only these three basic properties.}. The action (\ref{eqn:scalaraction}) is real, as seen by the following small calculation
\begin{gather*}
S_\star^\ast = -\frac{1}{2}\int \vol_\star^\ast\star\left( H_\star(d\Phi,d\Phi)^\ast + m^2 \Phi^\ast\star \Phi^\ast\right)\nonumber\\
\phantom{S_\star^\ast}{}
\overset{\vol_\star^\ast=\vol_\star,~\Phi^\ast=\Phi,~(\ref{eqn:hermitianproperties2})}{=}
-\frac{1}{2}\int \vol_\star\star \left(H_\star(d\Phi,d\Phi)+m^2 \Phi\star\Phi\right)\nonumber\\
\phantom{S_\star^\ast}{}
\overset{(\ref{eqn:gradedcyclicity})}{=} -\frac{1}{2}\int \left(H_\star (d\Phi,d\Phi)+m^2\Phi\star\Phi\right)\star\vol_\star=S_\star.
\end{gather*}
Interactions can be introduced to the free action (\ref{eqn:scalaraction}) by def\/ining
\[
S_{\star\text{int}} := -\int V_\star[\Phi]\star\vol_\star,\qquad V_\star[\Phi]^\ast =V_\star[\Phi],
\]
where $V_\star[\Phi]$ is a $\star$-deformed potential, e.g.~$V_\star[\Phi] = \frac{\lambda_4}{4!}\Phi\star\Phi\star\Phi\star\Phi$.

We calculate the equation of motion by demanding that the variation of the action vanishes, i.e.~$\delta S_\star =0$.
We def\/ine the  d'Alembert operator $\square_\star$ by
\[
\int \psi^\ast \star \square_\star[\varphi]\star \vol_\star := - \int H_\star(d\psi,d\varphi)\star \vol_\star ,
\]
for all $\psi,\varphi\in C^\infty(\mathcal{M})$ with
$\text{supp}(\psi)\cap\text{supp}(\varphi)$ compact.
The equation of motion obtained by $\delta S_\star=0$ is top-form valued and given by
\begin{gather}
\label{eqn:eomindep}
\tilde P_\star[\Phi] := \frac{1}{2}\Bigl(\square_\star[\Phi]\star\vol_\star + \vol_\star\star\bigl(\square_\star[\Phi^\ast] \bigr)^\ast - m^2 \Phi\star \vol_\star - m^2\vol_\star\star\Phi\Bigr)=0.
\end{gather}
Note that the equation of motion operator $\tilde P_\star$ is real.

Resembling standard formulations, we might extract the volume form to the right and def\/ine the scalar-valued
equation of motion operator $P_\star$ by $\tilde P_\star[\Phi]=:P_\star[\Phi]\star \vol_\star$.
It can be shown that the map
$C^\infty(\mathcal{M})\to\Omega^D(\mathcal{M}),~\varphi\mapsto \varphi\star\vol_\star$
is an isomorphism\footnote{The generalization to isomorphisms $\Omega^n\to\Omega^{D-n}$, and thus the construction of a NC Hodge operator,
is to our knowledge still an open problem, which, however, does not alter our construction of deformed scalar f\/ield theories.}. Its inverse is the right-extraction of $\vol_\star$, which
we have used to def\/ine the scalar-valued equation of motion operator $P_\star$.
As an aside, the operator $P_\star$ can be shown to be formally self-adjoint with respect to the deformed scalar product
\[
(\psi,\varphi)_\star :=\int \psi^\ast\star\varphi\star\vol_\star,
\]
i.e.~$(\psi,P_\star[\varphi])_\star = (P_\star[\psi],\varphi)$ for all $\psi,\varphi\in C^\infty(\mathcal{M})$
with compact overlap. This property is of particular importance for the construction of a QFT.

\subsection[Formulation in the coordinate and nice basis]{Formulation in the coordinate and {\it \textbf{nice}} basis}\label{sec:basis-formulation}
Let us now do the same construction as before in two dif\/ferent ``preferred'' bases of vector f\/ields. In physics, one
typically uses a basis $\partial_\mu$  of derivatives along some coordinates $x^\mu$. The inverse
metric f\/ield in this basis reads
\[
g^{-1} = \partial_\mu^\ast\otimes_\star g^{\mu\nu}\star\partial_\nu,\qquad\left( g^{\mu\nu}\right)^\ast=g^{\nu\mu}.
\]
Note that we used the conjugated basis vector f\/ield $\partial_\mu^\ast=-\partial_\mu$ in the left slot of the tensor product in order
 to avoid a minus sign in~(\ref{eqn:hermitian-in-coordinates}).
We furthermore use the dual basis $\widetilde{dx}^\mu$ def\/ined by $\langle \partial_\nu,\widetilde{dx}^\mu\rangle_\star=\delta_\nu^\mu$.
Note that, due to the deformed pairing, the commutative dual basis $dx^\mu$ def\/ined by $\langle \partial_\nu,dx^\mu\rangle=\delta_\nu^\mu$ is
not necessarily equal to the deformed dual basis $\widetilde{dx}^\mu$.
We express $d\Phi=:\widetilde{dx}^\mu\star\partial_{\star\mu}\Phi$ in terms of the NC basis.
The deformed derivatives $\partial_{\star\mu}$ are in general higher derivative operators obtained order by order in the deformation parameter
by solving $d\Phi=dx^\mu \partial_\mu\Phi \equiv \widetilde{dx}^\mu\star\partial_{\star\mu}\Phi$.
The hermitian structure in indices reads
\begin{gather}\label{eqn:hermitian-in-coordinates}
H_\star(d\Phi,d\Phi) \overset{(\ref{eqn:hermitianproperties3})}{=} \left(\partial_{\star\mu}\Phi\right)^\ast\star H_\star \big(\widetilde{dx}^\mu,\widetilde{dx}^\nu\big)\star \partial_{\star\nu}\Phi=\left(\partial_{\star\mu}\Phi\right)^\ast\star g^{\mu\nu}\star\partial_{\star\nu}\Phi.
\end{gather}
We f\/ind for the action (\ref{eqn:scalaraction})
\begin{gather}
\label{eqn:actioncoordinate}
S_\star = -\frac{1}{2}\int \left(\left(\partial_{\star\mu}\Phi\right)^\ast\star g^{\mu\nu}\star\partial_{\star\nu}\Phi + m^2 \Phi\star\Phi\right)\star\vol_\star,
\end{gather}
where we could also write $\vol_\star = \gamma\star dx^0\wedge_\star dx^1\wedge_\star \cdots\wedge_\star dx^{D-1}$, with some function
 $\gamma$ satisfying $\gamma=\sqrt{\vert g\vert}+\mathcal{O}(\lambda)$ to have a good classical limit.
Note that even though the action (\ref{eqn:actioncoordinate}) looks quite familiar and simple, it contains f\/irstly the metric in a nontrivial
basis (including $\star$-products) and the deformed derivatives $\partial_{\star\mu}$, which are higher dif\/ferential operators.
The equation of motion can be obtained using integration by parts order by order in the deformation parameter. We
do not derive it in detail, since~-- as we show now~-- there is a more convenient choice of basis, leading to a simpler form of the action.

As it was argued in \cite{Ohl:2009pv} and proven in detail in \cite{Aschieri:2009qh}, there is an almost everywhere def\/ined
basis of vector f\/ields $\lbrace e_a\rbrace$ and one-forms $\lbrace\theta^a\rbrace$ satisfying $[e_a,e_b]=0$, on which the RJS-twist~(\ref{eqn:rjstwist}) acts trivially. By acting trivially we mean that the
Lie derivatives along all $X_\alpha$ vanish,
i.e.~$\mathcal{L}_{X_\alpha} e_a =0$ and $\mathcal{L}_{X_\alpha}\theta^a=0$. This basis is called the {\it natural} or {\it nice} basis.
A similar notion of central bases, called ``frames'' or ``Stehbeins'', has already occurred in \cite{frame}.
For the {\it nice} basis NC duality
is equal to commutative duality, since $\langle e_a,\theta^b\rangle_\star = \langle e_a,\theta^b \rangle=\delta^b_a$.
We express the metric f\/ield in the {\it nice} basis $g^{-1}=e_a^\ast \otimes_\star g^{ab}\star e_b=e_a^\ast\otimes g^{ab}e_b$
and obtain that all $\star$-products drop out. Furthermore, we can write the derivative of $\Phi$ in this basis
and obtain $d\Phi=\theta^a\star e_a(\Phi)=\theta^a e_a(\Phi)$, where $e_a(\Phi)$
denotes the vector f\/ield action (Lie derivative) on $\Phi$. We also use the {\it nice} basis to write the volume form as
$\vol_\star = \gamma \star \theta^1\wedge_\star\theta^2\wedge_\star \cdots\wedge_\star\theta^D=\gamma   \theta^1\wedge\theta^2\wedge \cdots
\wedge\theta^D$. Note that the dif\/ferential form $\text{cnt}:=\theta^1\wedge\theta^2\wedge \cdots\wedge\theta^D$ is central,
i.e.~it $\star$-commutes with every function. Assuming the vector f\/ields $e_a$ to be real (this is typically the case),
the action (\ref{eqn:scalaraction}) reads
\[
S_\star = -\frac{1}{2}\int \left( e_a(\Phi)\star g^{ab}\star e_b(\Phi) + m^2 \Phi\star\Phi\right)\star\gamma \star\text{cnt}.
\]
The obvious advantage of the {\it nice} basis compared to (\ref{eqn:actioncoordinate}) is that no higher dif\/ferential ope\-ra\-tors
(such as $\partial_{\star\mu}$ above) occur. The equation of motion can be calculated by using
 graded cyclicity~(\ref{eqn:gradedcyclicity}), integration by parts and that the vector f\/ields $e_a$ act trivially on $\text{cnt}$. We obtain
\[
\tilde P_\star[\Phi]=\frac{1}{2} \left( e_a(g^{ab}\star e_b(\Phi)\star\gamma) + e_a(\gamma\star e_b(\Phi)\star g^{ba}) -m^2 (\Phi\star\gamma+\gamma\star\Phi)  \right)\star \text{cnt}=0.
\]
We again extract the volume form to the right and obtain the scalar-valued equation of motion
\begin{gather}
\label{eqn:eomnice}
P_\star[\Phi]=\frac{1}{2} \left( e_a(g^{ab}\star e_b(\Phi)\star\gamma) + e_a(\gamma\star e_b(\Phi)\star g^{ba}) -m^2 (\Phi\star\gamma+\gamma\star\Phi)  \right)\star \gamma^{-1\star}=0,\!\!
\end{gather}
where $\gamma^{-1\star}$ is the $\star$-inverse of $\gamma$ def\/ined by
 $\gamma\star\gamma^{-1\star}=\gamma^{-1\star}\star\gamma =1$.

\section[Examples~I: Deformed Klein-Gordon operators]{Examples~I: Deformed Klein--Gordon operators}\label{sec:examples1}

In this section we provide examples of deformed Klein--Gordon operators on NC
spacetimes, which solve the NC Einstein equations. For details on solutions of the NC
Einstein equations see \cite{Schupp:2009pt,Ohl:2009pv,Aschieri:2009qh} and also \cite{Asakawa:2009yb,Stern:2009id}
for related approaches. One of the main results of these papers is that the NC Einstein equations are solved by
the classical metric f\/ield, if the twist obeys certain properties. A suf\/f\/icient condition is given by
\[
\Theta^{\alpha\beta}X_\alpha\otimes X_\beta \in \Xi\otimes \mathfrak{g}+\mathfrak{g}\otimes\Xi,
\]
where $\mathfrak{g}$ is the Lie algebra of Killing vector f\/ields of the metric and $\Xi$ are general vector f\/ields.

\subsection{Deformed Minkowski spacetime}\label{subsec:minkex}

The simplest model we can consider is the Minkowski spacetime deformed by the Moyal--Weyl twist (\ref{eqn:moyaltwist}).
In this case the {\it nice} basis def\/ined above coincides with the coordinate basis in which the metric takes the form
$g^{-1}=\partial_\mu^\ast\otimes\eta^{\mu\nu}\partial_\nu$, where $\eta^{\mu\nu}=\text{diag}(-1,1,1,1)^{\mu\nu}$.
The volume form is given by $\vol_\star=dt\wedge dx^1\wedge dx^2\wedge dx^3=\text{cnt}$ and is central. In the language
 above, the function relating the volume form to the central form is $\gamma\equiv 1$.
Evaluating the equation of motion~(\ref{eqn:eomnice})
using $\lbrace e_a\rbrace=\lbrace\partial_\mu\rbrace$ we f\/ind
 \[
 P_\star[\Phi]=\eta^{\mu\nu}\partial_\mu\partial_\nu\Phi - m^2 \Phi=0.
 \]
 This result agrees with other approaches and shows that the free f\/ield on the Moyal--Weyl deformed Minkowski
 spacetime is not af\/fected by the deformation.

 Let us now consider a model leading to a deformed equation of motion. Consider the RJS twist (\ref{eqn:rjstwist})
 constructed from the vector f\/ields $X_1=\partial_t$ and $X_2=x^i\partial_i$, where $t$ is time and $x^i$ are spatial coordinates.
 This is an example of a Lie algebraic deformation $\starcom{t}{x^i}=i\lambda x^i$.
 One possible choice of a {\it nice} basis is given by
 \begin{gather}
 \label{eqn:nicekappa}
 e_1=\partial_t,\qquad e_2= r\partial_r,\qquad e_3=\partial_\zeta,\qquad e_4=\partial_\phi,
 \end{gather}
 where we introduced spherical coordinates $(r,\zeta,\phi)$. Note the additional $r$ in $e_2$ and that in spherical coordinates
 $X_2=r\partial_r$.
 We have $[e_a,e_b]=0$ and $\mathcal{L}_{X_\alpha}e_a=0$, thus $\lbrace e_a\rbrace$ indeed is a {\it nice} basis as def\/ined in Section \ref{sec:basis-formulation}.
 The dual basis is given by
 \begin{gather}
 \label{eqn:dualnicekappa}
 \theta^1=dt,\qquad \theta^2=\frac{dr}{r},\qquad \theta^3=d\zeta,\qquad \theta^4=d\phi,
 \end{gather}
 and is {\it nice}, too, i.e.~$\mathcal{L}_{X_\alpha}\theta^a=0$.
 In this basis the inverse metric f\/ield is given by $g^{-1}=e_a^\ast\otimes g^{ab}e_b$, where
  $g^{ab}=\text{diag}\left(-1,r^{-2},r^{-2},(r\sin\zeta)^{-2}\right)^{ab}$. We express the volume form as
   $\vol_\star=r^2\sin\zeta~ dt\wedge dr \wedge d\zeta\wedge d\phi=r^3\sin\zeta~\text{cnt}$, i.e.~$\gamma=r^3\sin\zeta$.
   Note the additional $r$ in $\gamma$ arising due to the form of $\theta^2$.
   Evaluating all $\star$-products, the equation of motion (\ref{eqn:eomnice}) reads
\begin{gather}
\label{eqn:eomminkowski}
P_\star[\Phi] = -\frac{1}{2}\big(1+e^{-i 3 \lambda\partial_t}\big) \big(\partial_t^2\Phi +m^2\Phi\big) +\frac{1}{2} \big(e^{i\lambda\partial_t}+e^{-i4\lambda\partial_t}\big)\triangle\Phi=0 ,
\end{gather}
where $\triangle=\partial^i\partial_i$ is the spatial Laplacian.
For deriving this equation one uses that $\sin\zeta$ is central and that for an arbitrary function $h\in C^\infty(\mathcal{M})$
and $n\in\mathbb{Z}$ the following identities hold true
\begin{gather}
\label{eqn:explicitstar}
(r^n)\star h = r^n e^{-\frac{in\lambda}{2}\partial_t}h,\qquad h\star (r^n)=r^n e^{\frac{in\lambda}{2}\partial_t}h.
\end{gather}
We thus obtain a nontrivial scalar f\/ield propagation on this deformed Minkowski spacetime.

\subsection{\label{subsec:frwex}Deformed FRW spacetime}
In \cite{Ohl:2009pv} we have studied NC (spatially f\/lat) FRW universes in presence of various twists.
Here we take two simple examples for illustration.
We again choose $X_1=\partial_t$ and $X_2=x^i\partial_i$,
leading to the same {\it nice} basis as above (\ref{eqn:nicekappa}), (\ref{eqn:dualnicekappa}). The inverse metric f\/ield in this basis
is given by $g^{-1}=e_a^\ast\otimes g^{ab}e_b$, where $g^{ab}=\text{diag}\left(-1,r^{-2} a(t)^{-2},r^{-2} a(t)^{-2},(r\sin\zeta)^{-2} a(t)^{-2}\right)^{ab}$
and $a(t)$ is the scale factor of the universe. Note again the additional $r^{-2}$ in the ``radial'' part of the metric, which arises
due to the choice of basis.
The volume form reads $\vol_\star = a(t)^3 r^2\sin\zeta~dt\wedge dr\wedge d\zeta \wedge d\phi=a(t)^3 r^3\sin\zeta~\text{cnt}$, i.e.~$\gamma=a(t)^3r^3\sin\zeta$.
In the following we restrict ourselves to a~universe which is a slice of de Sitter space, where $a(t)=e^{Ht}$ and $H$
is the Hubble constant. This drastically simplif\/ies the computation of the $\star$-products
 and thus of the equation of motion.
However, there are no obstructions in allowing a general scale factor $a(t)$.
After a straightforward calculation we obtain for the equation of motion (\ref{eqn:eomnice})
\begin{gather}
\label{eqn:waveFRW1}
P_\star[\Phi]=-\frac{1}{2} \big(1+ e^{-i3\lambda \mathcal{D}}\big) \big(\partial_t^2 + 3H\partial_t +m^2\big)\Phi +\frac{1}{2} \big(e^{i\lambda\mathcal{D}}+e^{-i4\lambda\mathcal{D}}\big) e^{-2Ht}\triangle\Phi=0,
\end{gather}
where $\mathcal{D}:=\partial_t-H r\partial_r$.
The following identities are used for deriving this expression
\begin{gather*}
h\star (a r)^n = (a r)^n e^{\frac{in\lambda}{2}\mathcal{D}}h,\qquad (a r)^n\star h = (a r)^n e^{-\frac{in\lambda}{2}\mathcal{D}}h,\qquad  \mathcal{D}(a r)^n = 0~,
\end{gather*}
which hold for all functions $h\in C^\infty(\mathcal{M})$ and $n\in\mathbb{Z}$ in case $a=e^{Ht}$.
Again, the free scalar f\/ield propagating on this spacetime is af\/fected by the~NC. Note that for $\lambda=0$
we obtain the usual equation of motion of a scalar f\/ield on de Sitter space and in the limit $H\to 0$ we obtain the equation
of motion on the deformed Minkowski spacetime~(\ref{eqn:eomminkowski}).

Next, we consider a model with nontrivial angle-time commutation relations. We choose $X_1=\partial_t$ and $X_2=L_3=\partial_\phi$, where $L_3$ denotes the angular momentum generator. As {\it nice} basis for this twist we can simply
use the spherical coordinate basis
\begin{gather*}
e_1=\partial_t,\qquad  e_2= \partial_r,\qquad e_3=\partial_\zeta,\qquad e_4=\partial_\phi,
\end{gather*}
and its dual
\begin{gather*}
 \theta^1=dt,\qquad \theta^2=dr,\qquad \theta^3=d\zeta,\qquad \theta^4=d\phi.
\end{gather*}
The inverse metric is $g^{-1}\!=\!e_a^\ast\otimes g^{ab} e_b$, where $g^{ab}\!=\!\text{diag}\left(-1,a(t)^{-2},r^{-2}a(t)^{-2},(r\sin\zeta)^{-2} a(t)^{-2}\right)^{ab}\!$.
The volume form is given by $\vol_\star=a(t)^3 r^2\sin\zeta~dt\wedge dr\wedge d\zeta \wedge d\phi = a(t)^3 r^2\sin\zeta~\text{cnt}$,
i.e.~$\gamma=a(t)^3 r^2\sin\zeta$. After a straightforward calculation we obtain for the equation of motion~(\ref{eqn:eomnice})
\begin{gather}
\label{eqn:waveFRW2}
P_\star[\Phi] = -\frac{1}{2}\big(1+e^{i3\lambda H \partial_\phi}\big) \big(\partial_t^2+3H\partial_t +m^2\big)\Phi +\frac{1}{2} \big(e^{-i\lambda H\partial_\phi}+e^{i4\lambda H\partial_\phi}\big) e^{-2Ht} \triangle \Phi =0.
\end{gather}
Again, the free f\/ield propagation is deformed.
The reason why these models lead to a deformed propagation, while the
Moyal--Weyl deformation of the Minkowski spacetime does not, is the fact that not all vector f\/ields occurring in the twist
are Killing vector f\/ields.

\subsection{Deformed Schwarzschild black hole}\label{subsec:bhex}

We brief\/ly study the equation of motion (\ref{eqn:eomnice}) on one of the NC Schwarzschild solutions
found in~\cite{Ohl:2009pv}. The choice of vector f\/ields $X_1=\partial_t$ and $X_2=x^i\partial_i$ is particularly interesting
 for the black hole, since the corresponding twist (\ref{eqn:rjstwist}) is then invariant under all classical symmetries,
 namely the spatial rotations and time translations. Furthermore, since $X_2$ is not a Killing vector f\/ield, we expect a deformed
  wave equation. We again use the {\it nice} basis (\ref{eqn:nicekappa}) and its dual (\ref{eqn:dualnicekappa}), and f\/ind
  for the inverse metric f\/ield in this basis $g^{ab}=\text{diag}\left( -Q(r)^{-1},Q(r) r^{-2},r^{-2},(r\sin\zeta)^{-2} \right)^{ab}$,
  where $Q(r)=1-\frac{r_s}{r}$ and $r_s$ is the Schwarzschild radius. The volume form is $\vol_\star = r^2\sin\zeta\,dt\wedge dr\wedge d\zeta\wedge d\phi = r^3\sin\zeta\,\text{cnt}$, i.e.~$\gamma=r^{3}\sin\zeta$. Evaluating the equation of motion
  (\ref{eqn:eomnice}) using (\ref{eqn:explicitstar}) we f\/ind
  \begin{gather}
  0=P_\star[\Phi] = -\frac{1}{2}\big(Q^{-1}\star\big(\partial_t^2\Phi\big) + \big(e^{-i3\lambda\partial_t}\partial_t^2\Phi\big)\star Q^{-1}\big)  -\frac{m^2}{2}\big(1+e^{-i3\lambda\partial_t}\big)\Phi\nonumber\\
  \phantom{0=}{}
  +\frac{1}{2r^2} \partial_r\left[ r^2\big(Q\star(\partial_r e^{i\lambda\partial_t}\Phi\big) + \big(\partial_r e^{-i4\lambda\partial_t}\Phi\big)\star Q\big)\right] +\frac{1}{2r^2} \big(e^{i\lambda\partial_t}+e^{-i4\lambda\partial_t}\big)\Delta_{S^2}\Phi
,\label{eqn:bhwave}
  \end{gather}
where $\Delta_{S^2}=\sin\zeta^{-1}\partial_\zeta\sin\zeta\partial_\zeta +\sin\zeta^{-2}\partial^2_\phi$ is the Laplacian on the unit
two-sphere $S^2$.
In addition to the exponentials of time derivatives, which we also found in the previous examples, there are the $\star$-products involving
either $Q(r)$ or $Q(r)^{-1}$.
While the former are easily evaluated, since $Q(r)$ is a sum of eigenfunctions of the dilation operator $r\partial_r$,
this is not the case for the latter.
Nevertheless, we can evaluate these products up to the desired order in the deformation
para\-me\-ter~$\lambda$ by using the explicit form of the $\star$-product~(\ref{eqn:rjsproduct}) and calculating
 the scale derivatives~$r\partial_r$ of~$Q^{-1}$.

\subsection[Deformed Randall-Sundrum spacetime]{Deformed Randall--Sundrum spacetime}

We now consider a deformation of the Randall--Sundrum (RS) spacetime, which is a slice of f\/ive-dimensional Anti de Sitter space AdS$_5$,
and is obtained as a solution of the classical Einstein equations with topology $\mathbb{R}^4\times S^1/\mathbb{Z}_2$ and two branes
localized at the orbifold f\/ixed points.
For an attempt to model building utilizing a NC RS spacetime and a discussion of the orbifold symmetry in the deformed case,
see~\cite{Ohl:2010bh}.
We employ 5D-coordinates
$x^M = (x^\mu,y)$, where $x^\mu$ are global coordinates on $\mathbb{R}^4$ and $y\in[0,\pi]$, in which the inverse metric is given by
\[
g^{-1} =\partial_M^\ast\otimes g^{MN}\partial_N= \partial_\mu^\ast \otimes e^{2kRy}\eta^{\mu\nu} \partial_\nu + \partial_y^\ast \otimes\frac{1}{R^2} \partial_y.
\]
Here $\eta^{\mu\nu}=\text{diag}(-1,1,1,1)^{\mu\nu}$ is the f\/lat metric, $R$ the radius of the extradimension and $k$ is related
to the curvature of the AdS space.
The deformation we consider is given by $2N$ commuting vector f\/ields $X_\alpha$ def\/ined as follows
\begin{gather}
\label{eqn:rsvectors}
X_{2j-1} = T_{2j-1}^{~\mu}\partial_\mu,\qquad X_{2j} = \vartheta(y) T_{2j}^{~\mu}\partial_\mu,\qquad\text{for} \ \ j=1,\dots,N,
\end{gather}
where $T_\alpha^{~\mu}$ are constant and $\vartheta(y)$ is a general smooth function. Note that $X_{2j-1}$
are Killing vector f\/ields of the RS metric for all $j$, and therefore the NC Einstein equations are solved for this model.
The RJS-twist~(\ref{eqn:rjstwist}) generated by the vector f\/ields~(\ref{eqn:rsvectors}) leads to the commutation relations
\[
\starcom{x^\mu}{x^\nu}=i\vartheta(y) \Theta^{\alpha\beta}T_\alpha^{~\mu}T_\beta^{~\nu}=:i \vartheta(y) \omega^{\mu\nu},\qquad \starcom{x^\mu}{y}=0.
\]
By a suitable choice of $T_\alpha^{~\mu}$ we can realize the most general constant antisymmetric matrix $\omega^{\mu\nu}$.

Let us now move on to f\/ield theory on that NC RS background. It turns out that, in contrast to previous examples,
the calculation of the action and equation of motion in the coordinate basis~(\ref{eqn:actioncoordinate}) is very simple. Thus, we do not use the {\it nice} basis here.
 The reason is that the metric coef\/f\/icients $g^{MN}$ are annihilated by all vector f\/ields $X_\alpha$,
 since $g^{MN}$ does not depend on~$x^\mu$. This leads to $g^{-1}=\partial_M^\ast\otimes g^{MN}\partial_N = \partial_M^\ast\otimes_\star g^{MN}\star \partial_N$. Furthermore, for the volume form $\vol_\star = e^{-4kRy} R \, dx^0\wedge dx^1\wedge dx^2\wedge dx^3\wedge dy$ we f\/ind $\mathcal{L}_{X_\alpha}\vol_\star=0$, for all $\alpha$.
 Using this and graded cyclicity of the integral,
 we obtain for the NC action (\ref{eqn:actioncoordinate})
 \begin{gather}
 \label{eqn:rsaction}
 S_\star= - \frac{1}{2}\int \left((\partial_{\star\mu} \Phi)^\ast\, e^{2kRy} \eta^{\mu\nu}\, \partial_{\star\nu}\Phi + (\partial_{\star y} \Phi)^\ast \frac{1}{R^2}\partial_{\star y}\Phi \right) e^{-4kRy} R\, d^4x\, dy.
 \end{gather}
 We have set the ``bulk mass'' $m^2=0$ for simplicity, and obtain ef\/fective mass terms via Kaluza-Klein reduction later.
The deformed derivatives are calculated by comparing both sides of
 $dx^M\, \partial_M\Phi = dx^M\star \partial_{\star M}\Phi$ and read
 \begin{gather}
 \label{eqn:rsdeformeddif}
 \partial_{\star\mu} = \partial_\mu,\qquad \partial_{\star y}=\partial_y +\frac{i\lambda}{2}\vartheta^\prime(y)\,\sum_{j=1}^N\big( T_{2j-1}^{~\mu} T_{2j}^{~\nu} \partial_\mu\partial_\nu\big)=:\partial_y +\frac{i\lambda}{2}\vartheta^\prime(y)\mathcal{T},
\end{gather}
where $\vartheta^\prime$ denotes the derivative of $\vartheta$.
Note that this result is exact, i.e.~it holds to all orders in~$\lambda$. Inserting (\ref{eqn:rsdeformeddif}) into (\ref{eqn:rsaction})
we obtain
\[
S_\star = -\frac{1}{2}\int \left(\partial_\mu\Phi \, e^{2kRy}\eta^{\mu\nu}\,\partial_\nu\Phi +\frac{1}{R^2}\partial_y\Phi\partial_y\Phi +\frac{\lambda^2}{4R^2} \vartheta^\prime(y)^2\,\mathcal{T}\Phi\, \mathcal{T}\Phi\right) e^{-4kRy} R\, d^4x\, dy~.
\]

We now turn to the Kaluza--Klein (KK) reduction of this action. We make the KK-ansatz
$\Phi(x^\mu,y) = \sum\limits_{n=0}^{\infty} \Phi_n(x^\mu)\, t_n(y)$, where $\Phi_n$ are the ef\/fective four-dimensional
f\/ields and $\lbrace t_n\rbrace$ is a~complete set of eigenfunctions of the mass operator
$\hat O=-R^{-2}\,e^{2kRy}\partial_ye^{-4kRy}\partial_y$ satisfying Neumann or Dirichlet boundary conditions.
The eigenfunctions $\lbrace t_n\rbrace$ are orthonormal
with respect to the standard scalar product, i.e.~$\int_0^\pi dy\,R\, e^{-2kRy}\, t_n t_m=\delta_{nm} $.
We obtain for the KK reduced action
\begin{gather}\label{eqn:KKactionRS}
S_\star = -\frac{1}{2}\sum\limits_{n=0}^{\infty }\int \left(\partial_\mu\Phi_n\, \eta^{\mu\nu}\,\partial_\nu\Phi_n +M_n^2\,\Phi_n^2 +\lambda^2 \sum\limits_{m=0}^{\infty}C_{nm}\,\mathcal{T}\Phi_n\, \mathcal{T}\Phi_m\right) \, d^4x,
\end{gather}
where the masses $M_n^2$ and the couplings $C_{nm}$ are given by
\[
\hat O t_n = M_n^2 t_n ,\qquad C_{nm} = \int\limits_{0}^\pi dy\, \frac{\vartheta^\prime(y)^2}{4R}   e^{-4kRy}  t_n(y) t_m(y).
\]
Thus, for the NC RS spacetime we f\/ind the standard ef\/fective 4D theory as obtained in the RS scenario,
but with additional Lorentz violating operators.
As an aside, note that in case
$\vartheta(y)\sim e^{kRy}$, which is one of the choices motivated in~\cite{Ohl:2010bh} from a dif\/ferent perspective,
the Lorentz violating operators are diagonal in the KK number.
The corresponding equations of motion for the $\Phi_n$ are derived easily from~(\ref{eqn:KKactionRS}),
so we do not provide them explicitly.

An interesting observation \cite{Schenkel:2010zi} is that we can,
as a special case of~(\ref{eqn:rsvectors}), obtain a deformation which yields a $z=2$~anisotropic propagator
(in the sense of~\cite{Horava:2009uw}) for the scalar f\/ield and does not af\/fect local potentials.
To this end, we specialize~(\ref{eqn:rsvectors}) to $N=3$ and $T_{2j-1}^{~\mu} = \delta_{j}^{\mu}$,
 $T_{2j}^{~\mu}=\delta_{j}^{\mu}$, resulting in $\mathcal{T}=\partial^i\partial_i=\triangle$. Choosing $\vartheta(y)$ such that
$C_{nm} = C_n \delta_{nm}$ is diagonal, we obtain propagator denominators of the form
\begin{gather}\label{eqn:z2-inv-propagator}
E^2- {\bf k}^2 -\lambda^2 C_{n} {\bf k}^4 -M_n^2,
\end{gather}
for all individual KK-modes. It is known that propagators of this kind improve the quantum behavior of
interacting  f\/ield theories, see e.g.~\cite{Horava:2009uw,Schenkel:2010zi}.
The problem of unitary ghosts, which typically arises in Lorentz invariant higher
derivative theories, is not present in our model since there time derivatives remain quadratic and only higher spatial derivatives occur.

\section{Perturbative approach to deformed Green's functions}\label{sec:greens}

In this section we provide an explicit formula for the retarded and advanced Green's functions
 of the deformed equation of motion  (\ref{eqn:eomindep}), see also (\ref{eqn:eomnice}) for an expression in
 terms of the {\it nice} basis. We always
 assume the classical spacetime obtained by setting $\lambda=0$ to be ``well-behaved''
  (mathematically speaking this means globally
  hyperbolic, time oriented and connected). Green's functions are not only of interest in classical f\/ield theory, but they also enter
  the def\/inition of the canonical commutator function of QFT, which in commutative QFT reads $[\Phi(x),\Phi(y)]=i \big(\tilde\Delta_+(x,y)-\tilde\Delta_-(x,y)\big)$, where $\tilde\Delta_{\pm}(x,y)$ denotes the retarded/advanced Green's function.

 To introduce a convenient notation, we consider a classical equation of motion opera\-tor~$P$, e.g.~a
 d'Alembert or Klein--Gordon operator. In physics literature, the Green's functions of $P$ are typically def\/ined as bi-distributions
 $\tilde\Delta_\pm(x,y)$ satisfying $P_x \tilde\Delta_\pm(x,y) = \delta(x,y)$ and $P_y\tilde\Delta_\pm(x,y)=\delta(x,y)$, where the labels $x$ and $y$
  denote the coordinates $P$ acts on and $\delta(x,y)$ is the (covariant) Dirac delta-function satisfying
  $\int \vol_y \delta(x,y) h(y) = h(x)$, for all test-functions $h\in C^\infty_0(\mathcal{M})$.
  Furthermore,  causality is used
  to distinguish between advanced and retarded. The retarded/advanced solution $\psi_\pm$ of the inhomogeneous
   problem $P[\psi_\pm]=\varphi$, where $\varphi$ denotes a source of compact support, is then obtained as the convolution
   of the Green's function and the source, i.e.~$\psi_\pm=\Delta_{\pm}[\varphi]:=\int \vol_y \tilde\Delta_{\pm}(x,y) \varphi(y)$.
In NC geometry it is convenient to work with the Green's operators $\Delta_{\pm}$ def\/ined above, instead of their integral kernels
 $\tilde\Delta_\pm(x,y)$.
   The def\/ining conditions $P_x \tilde\Delta_\pm(x,y) = \delta(x,y)$ and $P_y\tilde\Delta_\pm(x,y)=\delta(x,y)$ of the Green's
functions translate for the Green's operators to $P[\Delta_{\pm}[\varphi]]=\varphi$ and $\Delta_{\pm}[P[\varphi]]=\varphi$,
for all $\varphi\in C^\infty_0(\mathcal{M})$.

 In the NC case, we demand the deformed Green's operators $\Delta_{\star\pm}=\sum\limits_{n=0}^\infty\lambda^n\Delta_{(n)\pm}$
 to fulf\/il
\begin{gather*}
P_\star [\Delta_{\star\pm}[\varphi]] = \Delta_{\star\pm}[P_\star[\varphi]]=\varphi,
\end{gather*}
for all functions $\varphi$ with compact support.
We have proven in \cite{Ohl:2009qe} that the deformed Green's operators exist and also satisfy
the following causality condition
\begin{gather}
\label{eqn:green2}  \text{supp}(\Delta_{(n)\pm}[\varphi])\subseteq J_{\pm}(\text{supp}(\varphi)),
\end{gather}
for all $n$ and functions $\varphi$ with compact support, where $J_\pm(A)$ is the causal future/past of a~spacetime region $A$ w.r.t.~the classical metric $g\vert_{\lambda=0}$.
Note that (\ref{eqn:green2}) implies that the deformed propagation is compatible
with classical causality as determined by $g\vert_{\lambda=0}$.
For mathematical details we refer to the original work.

Additionally to the existence and uniqueness of the Green's operators, we have provided an explicit
formula for calculating the NC corrections
$\Delta_{(n)\pm}$ for $n>0$ in terms of the classical Green's operators $\Delta_{\pm}:=\Delta_{(0)\pm}$
and the deformed equation of motion operator $P_\star = \sum\limits_{n=0}^\infty\lambda^n P_{(n)}$.
Similar to standard perturbation theory, the NC corrections are given by composing the classical Green's operators
with the NC corrections of the equation of motion, precisely:
\begin{gather}
\label{eqn:explicitgreen}
\Delta_{(n)\pm} = \sum\limits_{k=1}^{n}\sum\limits_{j_1=1}^n \dots\sum\limits_{j_k=1}^n(-1)^k \delta_{j_1+\dots+j_k, n}
  \Delta_{\pm}\circ P_{(j_1)}\circ \Delta_\pm \circ P_{(j_2)} \circ \dots  \circ P_{(j_k)}\circ \Delta_{\pm},
\end{gather}
where $\delta_{n,m}$ is the Kronecker-delta and $\circ$ denotes the composition of operators,
i.e.~$(A\circ B)[\varphi]:=A[B[\varphi]]$ for two operators $A,B$ (maps from functions to functions).
Note that the composition of operators (\ref{eqn:explicitgreen}) might require an infrared
regularization in order to be mathematically well def\/ined. This can be achieved for example by introducing cutof\/f
functions $c_{(n)}\in C^\infty_0(\mathcal{M})$ of compact support and replacing $P_{(n)}$ by the regularized operators
$c_{(n)}\,P_{(n)}$, or by regularizing the twist (\ref{eqn:rjstwist}) by choosing vector f\/ields of compact support.
This is very similar to standard perturbative QFT, where all ``coupling constants'' have to be introduced as functions of
compact support in order to formulate a well def\/ined perturbation theory. After the calculation of physical observables,
one has to prove that the adiabatic limit $c_{(n)}\to 1$ exists, at least for all physical quantities.

\begin{figure}[h!]
\centerline{\includegraphics{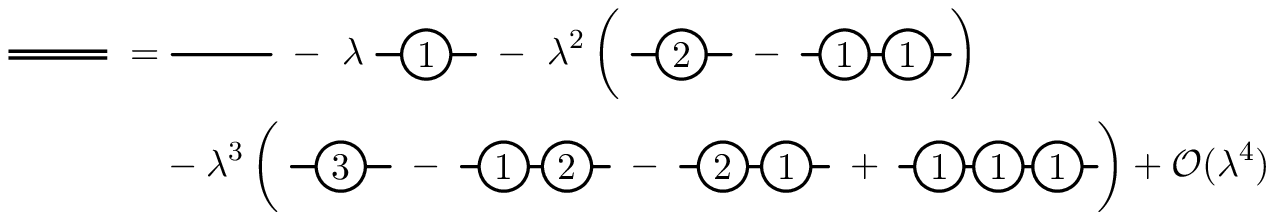}}
\caption{Diagrammatic representation of the NC retarded/advanced Green's operator (double line) in terms of the commutative retarded/advanced  Green's operator (single line) and the NC corrections to the equation of motion $P_{(n)}$ (vertices labeled by $n$).}\label{fig:perturbation}
\end{figure}

The expression (\ref{eqn:explicitgreen}) for the NC corrections to the Green's operators can be reformulated in a diagrammatic language as follows:
\begin{itemize}\itemsep=0pt
\item to the classical retarded/advanced Green's operator there corresponds a line
\item to each NC correction of the equation of motion operator $P_{(n)}$, $n>0$, there corresponds a~vertex labeled by $n$
\end{itemize}
The NC retarded/advanced Green's operator can then be represented graphically as shown in Fig.~\ref{fig:perturbation}.

\section{Examples~II: Deformed Green's functions}\label{sec:examples2}

In this section we derive the leading NC corrections to the Green's operators of
the NC Klein--Gordon operators studied in Section \ref{sec:examples1}.
For the deformed Randall--Sundrum spacetime, the derivation of the Green's functions
from (\ref{eqn:KKactionRS}) is straightforward, see (\ref{eqn:z2-inv-propagator}) for a particular choice of deformation.
We study the remaining examples in this section.
The focus is on the illustration of the formalism, but the results of this section can also be useful for
phenomenological studies of (quantum) f\/ield theories on curved NC spacetimes, e.g.~for NC cosmology or black hole physics.

\subsection{Deformed Minkowski spacetime}
We start with the simplest nontrivial example given by the equation of motion operator (\ref{eqn:eomminkowski})
on the $\kappa$-type deformed Minkowski spacetime. Even though this equation of motion operator can be diagonalized
using plane waves, we use the perturbative expansion (see Fig.~\ref{fig:perturbation}) in order to illustrate the
formalism.

The leading NC corrections of the equation of motion operator (\ref{eqn:eomminkowski}) are given by
\begin{gather*}
 P_{(1)}= \frac{3i}{2}\partial_t\left(\partial_t^2 -\triangle +m^2\right)=-\frac{3i}{2}\partial_t\circ P_{(0)},\\
 P_{(2)}=-\frac{9}{4} \partial_t^2\circ P_{(0)} -2 ~\partial_t^2\circ \triangle.
\end{gather*}
Note that the order $\lambda^1$ correction is imaginary. This {\it does not} violate the reality
of our f\/ield theory since the equation of motion operator resulting from the action $\tilde P_\star[\Phi] = P_\star[\Phi]\star\vol_\star$
is indeed real, but regarded as a top-form. It has been shown in~\cite{Ohl:2009qe} how to construct the space of real solutions of
such deformed wave operators.

We calculate the corrections to the Green's operators using Fig.~\ref{fig:perturbation}, and f\/ind
\begin{subequations}
\label{eqn:minkowskigreen}
\begin{gather}
 \Delta_{(1)\pm} = - \Delta_{\pm}\circ P_{(1)}\circ \Delta_{\pm} = \frac{3i}{2} \Delta_{\pm}\circ \partial_t\circ P_{(0)}\circ \Delta_{\pm}= \frac{3i}{2} \Delta_{\pm}\circ \partial_t,\\
\Delta_{(2)\pm}=2\Delta_{\pm}\circ\partial_t^2 \circ \triangle\circ\Delta_{\pm},
\end{gather}
\end{subequations}
where we have used $P_{(0)}\circ \Delta_{\pm}=\text{id}$, which results from the very def\/inition of the Green's operator.

Next, we extract the NC Green's functions $\tilde\Delta_{\star\pm}(x,y)$, i.e.~the integral kernels of the operators $\Delta_{\star\pm}$ def\/ined by
\[
 \Delta_{\star\pm}[\varphi](x)=:\int \tilde\Delta_{\star\pm}(x,y)\varphi(y)\vol_{y},
\]
for all functions $\varphi$ of compact support. Using (\ref{eqn:minkowskigreen}) and integration by parts we f\/ind
\begin{gather}
\label{eqn:explicitgreen2nd}
\tilde\Delta_{\star\pm}(x,y) =\tilde\Delta_{\pm}(x,y)-\frac{3i\lambda}{2}\partial_{t_y}\tilde\Delta_{\pm}(x,y) + 2 \lambda^2 \!\int\! \tilde\Delta_{\pm}(x,z)\partial_{t_z}^2\triangle_z\tilde\Delta_{\pm}(z,y) d^4z + \mathcal{O}(\lambda^3).\!\!
\end{gather}
The integral in the order-$\lambda^2$ part, $\tilde\Delta_{(2)\pm}$, can be evaluated explicitly using the momentum space representation
of the classical Green's functions\footnote{We use the standard convention $\tilde\Delta_\pm(x,y)=\lim\limits_{\epsilon\to0^+}\int\frac{d^4p}{(2\pi)^4} e^{-i p(x-y)} \left((p_0\pm i\epsilon)^2-\mathbf{p}^2-m^2\right)^{-1}$~.
}. It turns out that this integral does not require an infrared regularization and we obtain
\[
 \tilde \Delta_{(2)\pm}(x,y)=\mp \Theta\left(\pm t_z\right)\int\frac{d^3p}{(2\pi)^3} e^{-i\mathbf{p}\mathbf{z}} \mathbf{p}^2\left(t_z\cos(E_pt_z) + \frac{\sin(E_p t_z)}{E_p} \right),
\]
where $E_p=\sqrt{\mathbf{p}^2+m^2}$, $t_z=t_x-t_y$, $\mathbf{z}=\mathbf{x}-\mathbf{y}$ and $\Theta$ is the Heaviside step-function. For a~massless
f\/ield the remaining Fourier transformation is easily performed and one f\/inds that the support of the correction $\tilde\Delta_{(2)\pm}$
is, as expected, on the forward/backward lightcone. Since the explicit formula is not very instructive we do not include it here.

Note that the deformed Green's functions $\tilde\Delta_{\star\pm}(x,y)$ are not real.
This can be understood from the fact that the scalar-valued wave operator is not real, thus leading to complex Green's functions.
The sources to be considered physical
are those leading to real solutions of the inhomogeneous problem $P_{\star}[\psi]=\varphi$.
Multiplying both sides with the volume form $\vol_\star$ from the right we f\/ind $\tilde P_{\star}[\psi]=P_{\star}[\psi]\star\vol_\star =\varphi\star\vol_\star$.
Thus, the physical sources $\varphi$ have to obey the top-form reality condition
 $(\varphi\star\vol_\star)^\ast=\varphi\star\vol_\star$, which in general implies $\varphi^\ast\neq\varphi$ if the volume form is not central.

Applying the Green's operators to physical sources we f\/ind no nontrivial corrections at order~$\lambda^1$.
This is because a physical source $\varphi=\sum\lambda^n\varphi_{(n)}$ has to fulf\/il
$(\varphi\star\vol_\star)^\ast=\varphi\star\vol_\star$, which implies in our particular model
$\varphi_{(0)}^\ast=\varphi_{(0)}$ and $\mathfrak{Im}(\varphi_{(1)})=-\frac{3}{2}\partial_t\varphi_{(0)}$.
Thus, we obtain
\begin{gather*}
 \Delta_{\star\pm}[\varphi]
=\Delta_{\pm}\left[\varphi_{(0)}+\lambda  \varphi_{(1)} +\frac{3i\lambda}{2}\partial_t\varphi_{(0)}\right] + \mathcal{O}(\lambda^2)
=\Delta_{\pm}[\varphi_{(0)}+\lambda  \mathfrak{Re}(\varphi_{(1)})] + \mathcal{O}(\lambda^2).
\end{gather*}

\subsection{Deformed FRW spacetime}

We derive the second-order corrections to the Green's operators for the NC de Sitter universes discussed in Section \ref{subsec:frwex}.
Since the wave operators of both models are quite similar in their structure, see (\ref{eqn:waveFRW1}) and (\ref{eqn:waveFRW2}), we
derive the corrections for the f\/irst model and can obtain the corrections for the second model by replacing $\mathcal{D}\to-H\partial_\phi$.
Similar to the Minkowski model discussed before, the f\/irst nontrivial NC correction to the Green's operators acting on physical
sources is of order $\lambda^2$. The leading NC corrections of the wave operator (\ref{eqn:waveFRW1}) read
\begin{gather*}
 P_{(1)}=-\frac{3i}{2}\mathcal{D}\circ P_{(0)},\qquad
 P_{(2)}=-\frac{9}{4}\mathcal{D}^2\circ P_{(0)} -2 \mathcal{D}^2\circ \left(e^{-2Ht}\triangle\right).
\end{gather*}
Via Fig.~\ref{fig:perturbation} this leads to the following NC corrections to the Green's operators
\begin{gather*}
 \Delta_{(1)\pm}=\frac{3i}{2}\,\Delta_{\pm}\circ \mathcal{D},\qquad
\Delta_{(2)\pm} = 2\Delta_{\pm}\circ \mathcal{D}^2\circ\left(e^{-2Ht}\triangle\right)\circ \Delta_{\pm}.
\end{gather*}
The NC  integral kernel $\tilde\Delta_{\star\pm}(x,y)$ can be obtained by integration by parts and reads
\begin{gather*}
 \tilde\Delta_{\star\pm}(x,y) =
 \tilde\Delta_\pm(x,y)-\frac{3i\lambda}{2}\mathcal{D}_y\tilde\Delta_\pm(x,y)
 \nonumber\\
 \phantom{\tilde\Delta_{\star\pm}(x,y) =}{} + 2\lambda^2\int\tilde\Delta_{\pm}(x,z)\mathcal{D}^2_z\,e^{-2Ht_z}\triangle_z\tilde\Delta_{\pm}(z,y)
 \vol_z+\mathcal{O}(\lambda^3).
\end{gather*}
This shows that the corrections have a similar structure to the Minkowski case~(\ref{eqn:explicitgreen2nd}).
The explicit evaluation of the integral in the order-$\lambda^2$ part and the investigation of its IR regulator
(in)dependence is beyond the scope of this work.

\subsection{Deformed Schwarzschild black hole}
We derive for completeness the second-order corrections to the Green's operators for the NC Schwarz\-schild spacetime discussed in Section~\ref{subsec:bhex}.
The leading NC corrections of the equation of motion operator (\ref{eqn:bhwave}) read
\begin{gather*}
 P_{(1)}=-\frac{3i}{2}\,\partial_t\circ P_{(0)},\\
P_{(2)}=-\frac{9}{4}\partial_t^2\circ P_{(0)} - 2 \partial_t^2\circ \triangle_\mathrm{bh} + \mathfrak{B}_1 + \mathfrak{B}_2=:-\frac{9}{4}\partial_t^2\circ P_{(0)}+ \widehat P_{(2)},
\end{gather*}
where the spatial Laplacian $\triangle_\mathrm{bh}$ and the dif\/ferential operators $\mathfrak{B}_1$ and $\mathfrak{B}_2$ are def\/ined by
\begin{gather*}
\triangle_\mathrm{bh}[\varphi] :=\frac{1}{r^2}\partial_r\left(r^2 Q(r) \partial_r\varphi\right) + \frac{1}{r^2}\triangle_{S^2}\varphi,\\
\mathfrak{B}_1[\varphi]:=\frac{1}{8 Q(r)^3}  \frac{r_s}{r}\left(7-5\frac{r_s}{r}  \right) \partial_t^4\varphi,\\
\mathfrak{B}_2[\varphi]:=\frac{11 r_s}{8 r^2}\partial_r\left(r\partial_r\partial_t^2\varphi  \right).
\end{gather*}
Via Fig.~\ref{fig:perturbation} this leads to the following NC corrections to the Green's operators
\begin{gather*}
 \Delta_{(1)\pm}=\frac{3i}{2}\,\Delta_{\pm}\circ\partial_t,\\
\Delta_{(2)\pm} = \Delta_{\pm}\circ\left(2 \partial_t^2\circ \triangle_\mathrm{bh} -\mathfrak{B}_1 -\mathfrak{B}_2\right)\circ\Delta_\pm=-\Delta_{\pm}\circ\widehat P_{(2)}\circ\Delta_\pm.
\end{gather*}
The NC integral kernel $\tilde\Delta_{\star\pm}(x,y)$ can be obtained by integration by parts and reads
\begin{gather*}
\tilde\Delta_{\star\pm}(x,y) =  \tilde\Delta_\pm(x,y)-\frac{3i\lambda}{2}\partial_{t_y}\tilde\Delta_\pm(x,y)
-\lambda^2\int\tilde\Delta_{\pm}(x,z) \widehat P_{(2)z} \tilde\Delta_{\pm}(z,y) \vol_z+\mathcal{O}(\lambda^3).
\end{gather*}
Again, we do not evaluate the integral in the order-$\lambda^2$ part explicitly.
The calculation might be simplif\/ied drastically if one considers a two-dimensional reduction of the black hole by
only taking into account the isotropic modes (with spherical harmonic $Y_{00}(\zeta,\phi)$).

\section{Conclusions and outlook}\label{sec:conclusions}

In this article we have investigated classical scalar f\/ield theories on curved NC spacetimes,
with the NC deformations given by a large class of Drinfel'd twists.
Our models in particular include position dependent NC.
We have shown how to construct a deformed action for a~real scalar f\/ield and how to derive the corresponding equation
of motion in both, a geometric (global) and a coordinate-based (local) approach.
Subsequently, we have provided explicit examples of deformed Klein--Gordon operators on  NC Minkowski, de Sitter, Schwarzschild and
Randall--Sundrum spacetimes. Our deformed background spacetimes are chosen such that the NC Einstein equations of Wess  et al.\
\cite{Aschieri:2005yw,Aschieri:2005zs} are solved exactly.
We have then discussed the construction of the deformed Green's operators corresponding to the deformed wave operators and provided a diagrammatic
formalism for their perturbative calculation. The formalism has been applied to f\/ield theory on NC Minkowski, de Sitter and Schwarzschild spacetimes
in order to study the second-order correction to the advanced and retarded Green's functions.

This work is restricted to the level of classical f\/ield theory, since the construction of physical quantum states
in the formalism~\cite{Ohl:2009qe} has not been achieved yet. Nevertheless, the perturbative construction of Green's functions discussed
in the present paper can be used to construct the algebras of the corresponding QFT, since the canonical commutation relations are determined
by the Green's functions~\cite{Ohl:2009qe}. Once the construction of quantum states in our NC QFT is understood,
the results obtained here can be  applied in order to study NC ef\/fects in primordial power-spectra of scalar f\/ields
and NC ef\/fects in the vicinity of Schwarzschild black holes.
For the construction of quantum states the approach of \cite{Dappiaggi} might prove to be helpful.

\section*{Acknowledgements}
We thank Thorsten Ohl for comments and discussions on this work.
AS also thanks the Alessand\-ria Mathematical Physics Group, in particular Paolo Aschieri, and the
Vienna Mathematical Physics Group, in particular
Claudio Dappiaggi and Gandalf Lechner, for discussions and comments.
CFU is supported by the German National Academic Foundation
(Studienstiftung des deutschen Volkes).
AS and CFU are supported by Deutsche
Forschungsgemeinschaft through the Research Training Group GRK\,1147
\textit{Theoretical Astrophysics and Particle Physics}.

\pdfbookmark[1]{References}{ref}
 \LastPageEnding

\end{document}